\documentclass[a4paper]{article}

\usepackage[margin=1in]{geometry}

\usepackage[T1]{fontenc}
\newcommand{\changefont}[3]{
\fontfamily{#1} \fontseries{#2} \fontshape{#3} \selectfont}

\changefont{ptm}{m}{n}

\usepackage{setspace} 
 \usepackage{graphicx}    

\newcommand \be{\begin{equation}}
\newcommand \ee{\end{equation}}
\newcommand \ba{\begin{eqnarray}}
\newcommand \ea{\end{eqnarray}}

\def\bit{\begin{itemize}}
\def\eit{\end{itemize}}

\usepackage{amsfonts}
\usepackage{amssymb}
\usepackage{amsmath}
\usepackage{graphics}
\usepackage{mathrsfs}
\usepackage{color}

\newtheorem{theorem}{Theorem}[section]

\long\def\symbolfootnote[#1]#2{\begingroup%
\def\thefootnote{\fnsymbol{footnote}}\footnote[#1]{#2}\endgroup} 

\begin{document}

%

\begin{center}
\Large \textbf{Homoclinical Structure of Dynamic Equations on Time Scales}
\end{center}

\vspace{-0.3cm}
\begin{center}
\normalsize \textbf{Mehmet Onur Fen} \\
\vspace{0.2cm}
\textit{\textbf{\footnotesize Department of Mathematics, Middle East Technical University, 06800, Ankara, Turkey}} \\
\vspace{0.1cm}
\textit{\textbf{\footnotesize Department of Mathematics and Computer Science, \c{C}ankaya University, 06790, Ankara, Turkey}} \\
\vspace{0.1cm}
\textit{\textbf{\footnotesize E-mail: monur.fen@gmail.com}} \\
\vspace{0.1cm}
\end{center}

\vspace{0.3cm}

\begin{center}
\textbf{Abstract}
\end{center}

\noindent\ignorespaces

Homoclinic and heteroclinic motions in dynamics equations on time scales is investigated. The utilized time scale is a specific one such that it is a union of disjoint compact intervals.
A numerical example that supports the theoretical results is presented.

\vspace{0.2cm}
 
\noindent\ignorespaces \textbf{Keywords:} Dynamic equations on time scales; Homoclinic and heteroclinic motions; Stable and unstable sets

\section{Introduction}
 
Dynamic equations on time scales has become one of the attractive subjects among mathematicians beginning with the studies of S. Hilger (1988). The theory of time scales has many applications in various disciplines such as mechanics, electronics, neural networks, population models and economics (Bohner \& Peterson, 2001; Chen \& Song, 2013; Tisdell \& Zaidi, 2008; Zhang et al., 2010).

Homoclinic and heteroclinic motions play an important role in the theory of dynamical systems since they are concerned with the presence of chaos (Bertozzi, 1988; Chacon \& Bejarano, 1995; Gonchenko et al., 1996; Shil'nikov, 1967; Wiggins, 1988). The existence of a homoclinic solution was obtained as a limit of periodic solutions in a second order non-autonomous Hamiltonian system on time scales in the paper (Su \& Feng, 2014) by means of the mountain pass theorem and the standard minimizing argument. In the present paper, we consider differential equations on a specific time scale and show that perturbations through discrete equations give rise to the existence of homoclinic and heteroclinic motions in the system.

Let us denote by $\mathbb R$ and $\mathbb Z$ the sets of real numbers and integers, respectively. The basic concepts about differential equations on time scales are as follows (Bohner \& Peterson, 2001; Lakshmikantham et al. 1996; Lakshmikantham \& Vatsala, 2002; Lakshmikantham \& Devi, 2006). A time scale is a nonempty closed subset of $\mathbb{R}.$ On a time scale $\mathbb{T}$, the forward and backward jump operators are defined as $\sigma(t)=\inf\left\{s\in\mathbb{T}: s>t \right\}$ and $\rho(t)=\sup\left\{s\in\mathbb{T}: s<t\right\},$ respectively. We say that a point $t\in\mathbb{T}$ is right-scattered if $\sigma(t)>t$ and right-dense if $\sigma(t)=t$. Similarly, if $\rho(t)<t,$ then $t\in\mathbb{T}$ is called left-scattered, and otherwise it is called left-dense. Besides, a function $h:\mathbb{T} \times \mathbb{R}^n \to \mathbb{R}^n$ is called rd-continuous if it is continuous at each $(t,u) \in \mathbb{T} \times \mathbb{R}^n$ with right-dense $t,$ and the limits $\displaystyle \lim_{(r,\nu) \to (t^-,u)} h(r,\nu) =h(t-,u)$ and $\displaystyle \lim_{\nu \to u} h(t,\nu)=h(t,u)$ exist at each $(t,u)$ with left-dense $t.$ At a right-scattered point $t\in\mathbb{T},$ the $\Delta$-derivative of a continuous function $\vartheta$ is defined as $\vartheta^{\Delta}\left(t\right)=\displaystyle \frac{\vartheta\left(\sigma\left(t\right)\right)-\vartheta\left(t\right)}{\sigma\left(t\right)-t}.$ On the other hand, at a right-dense point $t,$ we have $\vartheta^{\Delta}\left(t\right)=\displaystyle \lim_{r \to t}\frac{\vartheta\left(t\right)-\vartheta\left(r\right)}{t-r}$ provided that the limit exists.

Let $\mathbb T_0$ be a time scale defined as $\mathbb T_0 = \bigcup_{k=-\infty}^{\infty} [\theta_{2k-1}, \theta_{2k}]$ in which $\left\{\theta_k\right\}_{k \in \mathbb Z}$ is a strictly increasing sequence of real numbers satisfying $\left|\theta_k\right| \to \infty$ as $\left|k\right|\to \infty.$  Suppose that there exist positive numbers $\bar{\kappa}$ and $\bar{\delta}$ such that $\theta_{2k}-\theta_{2k-1} \ge \bar{\kappa}$ and $\theta_{2k+1}-\theta_{2k} \le \bar{\delta}$ for all $k \in \mathbb Z.$

In the present study, we consider the following equation,
\begin{eqnarray} \label{main_eqn1}
y^{\Delta} (t)=Ay(t) + f(t,y(t)) + g(t,\zeta), \ t\in\mathbb{T}_0 ,
\end{eqnarray}
where $A$ is a constant $n\times n$ real valued matrix, the function $f:\mathbb T_0 \times \mathbb R^n \to \mathbb R^n$ is rd-continuous and the function $g(t,\zeta)$ is defined through the equation $g(t,\zeta)=\zeta_k$ for $t\in [\theta_{2k-1}, \theta_{2k}],$ $k\in\mathbb Z,$ such that $\zeta=\left\{\zeta_k\right\}_{k \in \mathbb Z}$ is a sequence generated by the map 
\begin{eqnarray}\label{timescale_map}
\zeta_{k+1}= F(\zeta_k), 
\end{eqnarray}
where $\zeta_0 \in \Lambda,$  $F:\Lambda \to \Lambda$ is a continuous function and $\Lambda$ is a bounded subset of $\mathbb R^n.$ We assume without loss of generality that $\theta_{-1}<0<\theta_0.$  Our aim in this paper is to show the existence of homoclinic and heteroclinic motions in the dynamics of system (\ref{main_eqn1}) in the case that the map (\ref{timescale_map}) admits homoclinic and heteroclinic orbits.

The presence of chaos in systems of the form (\ref{main_eqn1}) was considered in the paper (Akhmet \& Fen, 2015). The reduction technique to impulsive differential equations, which was presented by Akhmet and Turan (2006), was used in (Akhmet \& Fen, 2015) to investigate the existence of chaos. The reader is referred to (Akhmet, 2009a, 2009b; Akhmet \& Fen, 2012, 2013a, 2013b, 2014a, 2014b, 2016) for other chaos generation techniques in differential equations. In the literature, the generation of homoclinic and heteroclinic motions on the basis of functional spaces was first demonstrated in (Akhmet, 2008, 2010a). It was rigorously proved in (Akhmet, 2010a) that the chaotic attractors of relay systems contain homoclinic motions. By means of the structure of moments of impulses, similar results for differential equations with impacts were obtained in the study (Akhmet, 2008). On the other hand, taking advantage of perturbations, the presence of homoclinic and heteroclinic motions in impulsive differential equations was demonstrated in (Fen \& Tokmak Fen, accepted), and an illustrative example concerning Duffing equations with impacts was provided. However, dynamic equations on time scales were not investigated in these papers. The novelty of the present study is the theoretical discussion of homoclinic and heteroclinic motions in dynamic equations on time scales. The definitions for the stable and unstable sets as well as homoclinic and heteroclinic motions on time scales are given on the basis of functional spaces. Simulations that support the theoretical results are also provided.

\section{Preliminaries} \label{prelim}

Throughout the paper, the usual Euclidean norm for vectors and the norm induced by the Euclidean norm for square matrices (Horn \& Johnson, 1992) will be utilized.  

Let us take into account the function $\psi: \mathbb T'_0 \to \mathbb R$ defined as (Akhmet \& Turan, 2006),   
\begin{eqnarray*} 
\displaystyle \psi(t)=\left\{\begin{array}{ll} \displaystyle t-\sum_{0<\theta_{2k}<t} \delta_k,    &   t  \ge 0, \\
                                                 \displaystyle t+\sum_{t\le\theta_{2k}<0} \delta_k,    &   t < 0, 
\end{array} \right.
\end{eqnarray*}
where $\delta_k=\theta_{2k+1}-\theta_{2k},$ $k \in \mathbb Z,$ and $\mathbb T'_0 = \mathbb T_0 \setminus \bigcup_{k =-\infty}^{\infty} \left\{\theta_{2k-1} \right\},$ and denote by $X(s,r)$ the transition matrix of the linear homogeneous impulsive system
\begin{eqnarray*} 
\begin{array}{l}
x'(s)=Ax(s), ~s \neq s_k,\\
\Delta x|_{s=s_k} =  \delta_k A x(s_k),
\end{array}
\end{eqnarray*} 
where $s_k=\psi(\theta_{2k}),$ $k \in \mathbb Z,$ $\displaystyle \Delta x |_{s=s_k}=x(s_k+)-x(s_k)$ and $\displaystyle x(s_k+)=\lim_{s\to s_k^+} x(s).$

The following conditions are required.
\begin{enumerate}
\item[\textbf{(C1)}] $\det(I + \delta_kA) \neq 0$ for all $k\in \mathbb Z,$ where $I$ is the $n \times n$ identity matrix;
\item[\textbf{(C2)}] There exist positive numbers $N$ and $\lambda$ such that $\left\|X(s,r)\right\| \le Ne^{-\lambda (s-r)}$ for $s\ge r;$ 
\item[\textbf{(C3)}] There exist positive numbers $M_f$ and $M_F$ such that $\displaystyle \sup_{t\in \mathbb T_0, ~y\in \mathbb R^n} \left\|f(t,y)\right\| \le M_f$ and $\displaystyle \sup_{\upsilon \in \Lambda} \left\|F(\upsilon)\right\|\le M_F;$ 
\item[\textbf{(C4)}] There exists a positive number $L_f$ such that $\left\|f(t,y_1)-f(t,y_2)\right\| \le L_f \left\|y_1-y_2\right\|$ for all $t\in \mathbb T_0,$ $y_1,y_2\in \mathbb R^n.$ 
\item[\textbf{(C5)}] $\displaystyle N L_f\left( \frac{1}{\lambda} + \frac{\bar{\delta}}{1-e^{-\lambda \bar{\kappa}}} \right)<1;$ 
\item[\textbf{(C6)}] $\displaystyle -\lambda+NL_f+ \frac{1}{\bar{\kappa}} \ln \left(1+NL_f\bar{\delta}\right)<0.$ 
\end{enumerate}

If the conditions $(C1)-(C5)$ are valid, then for a fixed solution $\zeta=\left\{\zeta_k\right\}_{k\in \mathbb Z}$ of (\ref{timescale_map}) there exists a unique solution $\varphi_{\zeta}(t)$ of (\ref{main_eqn1}) which is bounded on $\mathbb T_0$ such that $$\displaystyle \sup_{t\in\mathbb T_0} \left\|\varphi_{\zeta}(t)\right\| \le N(M_f+M_F) \Big(\displaystyle\frac{1}{\lambda}+\displaystyle\frac{\bar{\delta}}{1-e^{-\lambda \bar{\kappa}}}\Big).$$ Moreover, if $(C6)$ additionally holds,  then by using the Gronwall's Lemma for piecewise continuous functions (Akhmet, 2010b; Samoilenko \& Perestyuk, 1995) one can confirm that the bounded solution $\varphi_{\zeta}(t)$ attracts all other solutions of (\ref{main_eqn1}) for a fixed sequence $\zeta,$ that is, $\left\|y(t)-\varphi_{\zeta}(t)\right\|\to 0$ as $t\to \infty,$ $t\in\mathbb T_0,$ where $y(t),$ $y(t^0)=y_0,$ is a solution of (\ref{main_eqn1}) for some $t^0 \in \mathbb T_0$ and $y_0\in \mathbb R^n.$

In the next section, we will investigate the existence of homoclinic and heteroclinic motions in the dynamics of (\ref{main_eqn1}).

\section{Homoclinic and Heteroclinic Motions}

Denote by $\mathcal{H}$ the set of all sequences  $\zeta=\left\{\zeta_k\right\}_{k\in \mathbb Z}$ obtained by the discrete map (\ref{timescale_map}). The following definitions are adapted from the papers (Akhmet, 2008, 2010a).

The stable set of a sequence $\zeta \in \mathcal{H}$ is defined as 
\begin{eqnarray*}  
W^s(\zeta)= \left\{ \eta=\left\{\eta_k\right\}_{k\in \mathbb Z} \in \mathcal{H} \ : \ \left\|\eta_k-\zeta_k\right\|\to 0 ~\textrm{as}~ k\to  \infty  \right\},
\end{eqnarray*}
and the unstable set of $\zeta$ is 
\begin{eqnarray*}  
W^u(\zeta)= \left\{ \eta=\left\{\eta_k\right\}_{k\in \mathbb Z} \in \mathcal{H} \ : \ \left\|\eta_k-\zeta_k\right\|\to 0 ~\textrm{as}~ k\to  -\infty  \right\}.
\end{eqnarray*}
The set $\mathcal{H}$ is called hyperbolic if for each $\zeta \in \mathcal{H}$ the stable and unstable sets of $\zeta$ contain at least one element different from $\zeta.$ A sequence $\eta \in \mathcal{H}$ is homoclinic to another sequence $\zeta \in \mathcal{H}$ if $\eta \in W^s(\zeta) \cap W^u(\zeta).$ Moreover, $\eta \in \mathcal{H}$ is heteroclinic to the sequences $\zeta^1,$ $\zeta^2 \in \mathcal{H},$ $\eta \neq \zeta^1,$ $\eta \neq \zeta^2,$ provided that $\eta \in W^s(\zeta^1) \cap W^u(\zeta^2).$

Suppose that $\mathcal{D}$ is the set consisting of all solutions $\varphi_{\zeta}(t),$ $\zeta \in \mathcal{H},$ of (\ref{main_eqn1}), which are bounded on $\mathbb T_0.$ 
A solution $\varphi_{\eta}(t) \in \mathcal{D}$ belongs to the stable set $W^s(\varphi_{\zeta}(t))$ of  $\varphi_{\zeta}(t) \in \mathcal{D}$ if $\left\|\varphi_{\eta}(t)-\varphi_{\zeta}(t)\right\|\to 0$ as $t\to \infty,$ $t\in \mathbb T_0.$ Besides, $\varphi_{\eta}(t)$ is an element of the unstable set $W^u(\varphi_{\zeta}(t))$ of $\varphi_{\zeta}(t)$ provided that $\left\|\varphi_{\eta}(t)-\varphi_{\zeta}(t)\right\|\to 0$ as $t\to -\infty,$ $t \in \mathbb T_0.$ 

We say that $\mathcal{D}$ is hyperbolic if for each  $\varphi_{\zeta}(t) \in \mathcal{D}$ the sets $W^s(\varphi_{\zeta}(t))$ and $W^u(\varphi_{\zeta}(t))$ contain at least one element different from $\varphi_{\zeta}(t).$ A solution $\varphi_{\eta}(t)\in \mathcal{D}$ is homoclinic to another solution $\varphi_{\zeta}(t) \in \mathcal{D}$ if $\varphi_{\eta}(t) \in W^s(\varphi_{\zeta}(t)) \cap W^u(\varphi_{\zeta}(t)),$ and $\varphi_{\eta}(t)\in \mathcal{D}$ is heteroclinic to the bounded solutions $\varphi_{\zeta^1}(t),$ $\varphi_{\zeta^2}(t)\in \mathcal{D},$ $\varphi_{\eta}(t) \neq \varphi_{\zeta^1}(t),$ $\varphi_{\eta}(t) \neq \varphi_{\zeta^2}(t),$ if $\varphi_{\eta}(t) \in W^s(\varphi_{\zeta^1}(t)) \cap W^u(\varphi_{\zeta^2}(t)).$

The following theorem can be proved by using the reduction technique to impulsive differential equations, which was introduced in (Akhmet \& Turan, 2006).

\begin{theorem}\label{main_theorem}
Under the conditions $(C1)-(C6),$ the following assertions are valid:
\begin{enumerate}
\item[(i)] If $\eta \in \mathcal{H}$ is homoclinic to $\zeta \in \mathcal{H},$ then $\varphi_{\eta}(t) \in \mathcal{D}$ is homoclinic to $\varphi_{\zeta}(t) \in \mathcal{D};$
\item[(ii)] If $\eta \in \mathcal{H}$ is heteroclinic to $\zeta^1,$ $\zeta^2 \in \mathcal{H},$ then $\varphi_{\eta}(t) \in \mathcal{D}$ is heteroclinic to $\varphi_{\zeta^1}(t),$ $\varphi_{\zeta^2}(t) \in \mathcal{D};$
\item[(iii)] If $\mathcal{H}$ is hyperbolic, then the same is true for $\mathcal{D}.$
\end{enumerate}
\end{theorem}

The next section is devoted to a numerical example, which supports the results of Theorem \ref{main_theorem}.

\section{An Example} \label{example_sec}

Consider the following system,
\begin{eqnarray} \label{example_1}
\begin{array}{l}
y_1^{\Delta}(t) = \displaystyle - (11/60) y_1(t) + (1/30) y_2(t) + 0.05\sin\left(\pi t / 4\right) + g(t, \zeta), \\   
y_2^{\Delta}(t) = \displaystyle - (1/20) y_1(t) - (1/15) y_2(t) + 0.003 \arctan(y_1(t)),
\end{array}
\end{eqnarray}
where $t$ belongs to the time scale $\mathbb T_0= \bigcup_{k=-\infty}^{\infty} [\theta_{2k-1}, \theta_{2k}],$ $\theta_k=4k + (5+(-1)^k)/2$ and $g(t,\zeta)=\zeta_k$ for $t\in [\theta_{2k-1}, \theta_{2k}],$ $k\in\mathbb Z,$ such that $\zeta=\left\{\zeta_k\right\}_{k \in \mathbb Z},$ $\zeta_0 \in [0,1],$ is a sequence generated by the logistic map
\begin{eqnarray}\label{logistic_map}
\zeta_{k+1} = F(\zeta_k), 
\end{eqnarray}
where $F(\upsilon)= 3.9 \upsilon (1-\upsilon).$ It is worth noting that the map (\ref{logistic_map}) is chaotic in the sense of Li-Yorke (Li \& Yorke, 1975), and the unit interval $[0,1]$ is invariant under the iterations of the map (Hale \& Ko\c{c}ak, 1991).

System (\ref{example_1}) is in the form of (\ref{main_eqn1}), where 
$$ \displaystyle A=\left(
\begin {array}{ccc}
- \displaystyle \frac{11}{60} & \displaystyle \frac{1}{30}\\
\noalign{\medskip}
- \displaystyle \frac{1}{20} &- \displaystyle \frac{1}{15}
\end {array}
\right),  \  \ 
f(t,y_1,y_2)=\left(
\begin {array}{ccc}
0.05 \displaystyle \sin \left(\frac{\pi}{4}t\right)\\
\noalign{\medskip}
0.003 \arctan (y_1)  
\end {array}
\right).$$  
The conditions $(C1)-(C6)$ are valid for system (\ref{example_1}) with $N=2.62,$ $\lambda=1/12,$ $M_f=0.0503,$ $L_f=0.003,$ $\bar{\kappa}=5$ and $\bar{\delta}=3.$

\begin{figure}[ht] 
\centering
\includegraphics[width=11.2cm]{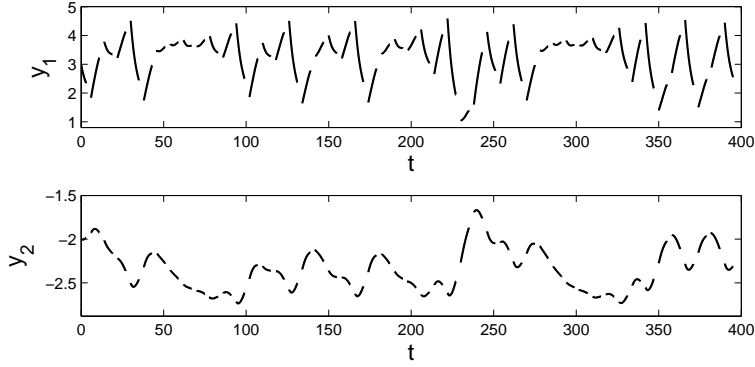}
\caption{The solution of (\ref{example_1}) corresponding to the initial data $y_1(0)=3,$ $y_2(0)=-2,$ $\zeta_0=0.292.$ The figure confirms the presence of chaos in system (\ref{example_1}).}
\label{fig1}
\end{figure}

Since $t+8 \in \mathbb T_0$ whenever $t\in \mathbb T_0$ and $f(t+8,y_1,y_2)=f(t,y_1,y_2)$ for all $t\in \mathbb T_0,$ $y_1,$ $y_2 \in \mathbb R,$ system (\ref{example_1}) is Li-Yorke chaotic according to the results of (Akhmet \& Fen, 2015). Moreover, one can verify that if $\zeta=\left\{\zeta_k\right\}_{k \in \mathbb Z}$ is a $p$-periodic solution of (\ref{logistic_map}) for some natural number $p,$ then the corresponding bounded solution $\varphi_{\zeta}(t)$ of (\ref{example_1}) is $8p$-periodic. The $y_1$ and $y_2$ coordinates of the solution of (\ref{example_1}) with $y_1(0)=3,$ $y_2(0)=-2,$ $\zeta_0=0.292$ is shown in Figure \ref{fig1}. It is seen in Figure \ref{fig1} that the solution behaves chaotically. 

\begin{figure}[ht] 
\centering
\includegraphics[width=11.0cm]{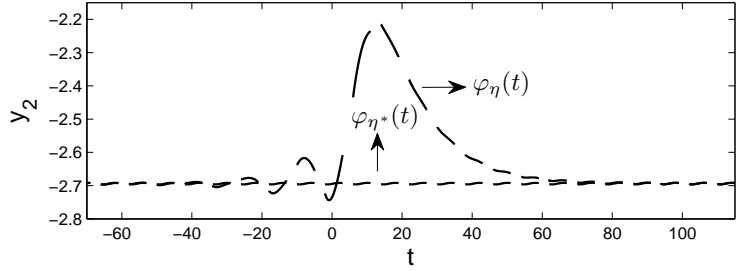}
\caption{The homoclinic solution of (\ref{example_1}). The simulation demonstrates that $\left\|\varphi_{\eta}(t)-\varphi_{\eta^*}(t)\right\| \to 0$ as $t \to \pm \infty,$ $t\in \mathbb T_0.$}
\label{fig2}
\end{figure}

Now, we will demonstrate that homoclinic and heteroclinic motions take place in the chaotic dynamics of (\ref{example_1}). Consider the function $\gamma_1(\upsilon)=\displaystyle \frac{1}{2} \left( 1+\sqrt{1-\frac{4\upsilon}{3.9}} \right).$ It was mentioned in (Avrutin et al., 2015) that the orbit $$\eta=\left\{\ldots, \gamma^3_1(\eta_0), \gamma_1^2(\eta_0), \gamma_1(\eta_0), \eta_0, F(\eta_0), F^2(\eta_0), F^3(\eta_0), \ldots \right\},$$ where $\eta_0=1/3.9,$ is homoclinic to the fixed point $\eta^{*}=2.9/3.9$ of (\ref{logistic_map}). Denote by $\varphi_{\eta}(t)$ and $\varphi_{\eta^*}(t)$ the bounded solutions of (\ref{example_1}) corresponding to $\eta$ and $\eta^*,$ respectively. Theorem \ref{main_theorem} implies that $\varphi_{\eta}(t)$ is homoclinic to $\varphi_{\eta^*}(t).$ Notice that the bounded solution $\varphi_{\eta^*}(t)$ is $8$-periodic. We depict in Figure \ref{fig2} the $y_2$ coordinates of $\varphi_{\eta}(t)$ and $\varphi_{\eta^*}(t).$ Figure \ref{fig2} supports the result of Theorem \ref{main_theorem} such that $\varphi_{\eta}(t)$ is homoclinic to $\varphi_{\eta^*}(t).$

\begin{figure}[ht] 
\centering
\includegraphics[width=11.0cm]{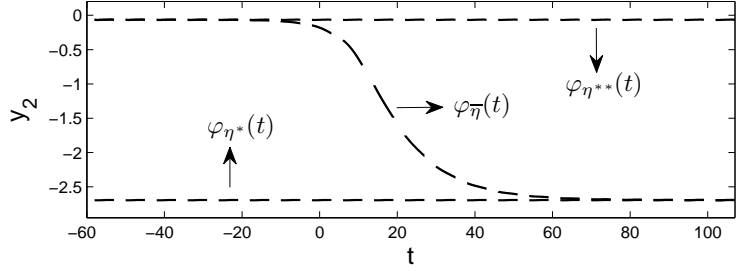}
\caption{The heteroclinic solution of (\ref{example_1}). It is observable that the bounded solution $\varphi_{\overline{\eta}}(t)$ is heteroclinic to the $8$-periodic solutions $\varphi_{\eta^*}(t),$ $\varphi_{\eta^{**}}(t).$}
\label{fig3}
\end{figure}

On the other hand, the orbit $$\overline{\eta}=\left\{\ldots, \gamma^3_2(\overline{\eta}_0), \gamma_2^2(\overline{\eta}_0), \gamma_2(\overline{\eta}_0), \overline{\eta}_0, F(\overline{\eta}_0), F^2(\overline{\eta}_0), F^3(\overline{\eta}_0), \ldots \right\},$$ where the function $\gamma_2$ is defined as $\gamma_2(\upsilon)=\displaystyle \frac{1}{2} \left( 1-\sqrt{1-\frac{4\upsilon}{3.9}} \right)$ and $\overline{\eta}_0=1/3.9,$ is heteroclinic to the fixed points $\eta^*=2.9/3.9$ and $\eta^{**}=0$ of (\ref{logistic_map}) according to the results of (Avrutin et al., 2015). Suppose that $\varphi_{\overline{\eta}}(t),$ $\varphi_{\eta^*}(t)$ and $\varphi_{\eta^{**}}(t)$ are the bounded solutions of (\ref{example_1}) corresponding to the sequences $\overline{\eta},$ $\eta^*$ and $\eta^{**},$ respectively. One can confirm by using Theorem \ref{main_theorem} that the solution $\varphi_{\overline{\eta}}(t)$ is heteroclinic to $\varphi_{\eta^*}(t),$ $\varphi_{\eta^{**}}(t).$ The $y_2$ coordinates of $\varphi_{\overline{\eta}}(t),$ $\varphi_{\eta^*}(t)$ and $\varphi_{\eta^{**}}(t)$ are represented in Figure \ref{fig3}, which reveals that $\left\|\varphi_{\overline{\eta}}(t)-\varphi_{\eta^*}(t)\right\| \to 0$ as $t \to \infty,$ $t \in \mathbb T_0$ and $\left\|\varphi_{\overline{\eta}}(t)-\varphi_{\eta^{**}}(t)\right\| \to 0$ as $t \to -\infty,$ $t \in \mathbb T_0.$

\section{Conclusion}

The present study is devoted to the existence of homoclinic and heteroclinic motions in dynamic equations on time scales. Sufficient conditions that guarantee the existence of homoclinic and heteroclinic motions are provided, and the definitions of stable and unstable sets are given on the basis of functional spaces. The equation under investigation can be considered as a hybrid system since it combines the dynamics of the discrete map with the continuous dynamics of the equation on time scales. The presented example and simulations reveal the applicability of our theoretical results.

\section*{Acknowledgments}

This work is supported by the 2219 scholarship programme of T\"{U}B\.{I}TAK, the Scientific and Technological Research Council of Turkey.


\end{document}